# Missing Mass in Collisional Debris from Galaxies

Frédéric Bournaud[1*], Pierre-Alain Duc[1], Elias Brinks[2], Médéric Boquien[1], Philippe Amram[3], Ute Lisenfeld[4,5], Bärbel S. Koribalski[6], Fabian Walter[7] & Vassilis Charmandaris[8,9,10]

[1]Laboratoire AIM, CEA/DSM - CNRS - Université Paris Diderot, DAPNIA/Service d'Astrophysique, CEA/Saclay, F-91191 Gif-sur-Yvette Cedex, France.
[2]Centre for Astrophysics Research, University of Hertfordshire, College Lane, Hatfield, AL10 9AB, UK.
[3]Observatoire Astronomique Marseille-Provence, LAM-UMR 6110, 2 place Le Verrier, F-13248 Marseille Cedex 4, France.
[4]Departamento Física Teórica y del Cosmos, Universidad de Granada, Spain.
[5]Instituto de Astrofísica de Andalucía (CSIC), PO Box 3004, 18080 Granada, Spain.
[6]CSIRO, Australia Telescope National Facility (ATNF), PO Box 76, Epping NSW 1710, Australia.
[7]Max Planck Institut für Astronomie, Königstuhl 17, 69117 Heidelberg, Germany.
[8]Department of Physics, University of Crete, GR-71003 Heraklion, Greece.
[9]IESL/Foundation for Research and Technology, Hellas, GR-71110, Heraklion, Greece.
[10]Observatoire de Paris, F-75014, Paris, France.
[*]To whom correspondence should be addressed. E-mail: frederic.bournaud@cea.fr

**Recycled dwarf galaxies can form in the collisional debris of massive galaxies. Theoretical models predict that, contrary to classical galaxies, they should be free of non-baryonic Dark Matter. Analyzing the observed gas kinematics of such recycled galaxies with the help of a numerical model, we demonstrate that they do contain a massive dark component amounting to about twice the visible matter. Staying within the standard cosmological framework, this result most likely indicates the presence of large amounts of unseen, presumably cold, molecular gas. This additional mass should be present in the disks of their progenitor spiral galaxies, accounting for a significant part of the so-called missing baryons.**

When galaxies collide, gravitational forces cause the expulsion of material from their disks into the intergalactic medium. In this debris, dense self-gravitating structures form. They can reach masses typical of those of dwarf galaxies, show ordered rotation and active star formation[1-8], hence deserve to be considered galaxies in their own right, albeit "recycled" ones. Whether these recycled dwarf galaxies contain dark matter can put strong constraints on the nature and distribution of this enigmatic constituent of the Universe. Indeed, standard theory[9-11] predicts that they differ from classical galaxies by being nearly free of non-baryonic dark matter[5,7,12]. According to the widely accepted ΛCDM (Cold Dark Matter with cosmological constant) model[13], the matter density of the Universe is dominated by non-baryonic dark matter. This matter is expected to surround galaxies in the form of large halos supported by random motions[9]. Recycled galaxies are expected to have little, if any, dark matter of this type, because only material from rotating disks is involved in the galactic recycling process. In addition to non-baryonic dark matter, part of the baryonic component is "dark" as well, known to exist in the early Universe[14], but hard or impossible to detect locally today. It has been speculated to be cold gas[15,16] but is most widely thought to reside in a diffuse warm-hot intergalactic medium (WHIM) surrounding galaxies[10,11], which cannot be substantially accumulated in collisional debris. Hence, recycled dwarf galaxies are predicted by conventional views to be mostly free of both baryonic and non-baryonic dark matter. We put these views to the test, measuring the mass of three galaxies formed in the collisional debris around galaxy NGC5291[17,18].

The galaxy NGC5291 is surrounded by a large, gas-rich ring of collisional debris[17]. In several places gas has gathered into self-gravitating, rotating dwarf galaxies where new stars form (Fig. 1). We study the kinematics of atomic hydrogen in the ring via its 21-cm emission line, using the NRAO[19] Very Large Array (VLA) interferometer in a high-resolution configuration. We estimate the mass actually present in the dwarf galaxies and compare this to their visible mass[6,18,20,21]. We use N-body simulations that model the gravitational dynamics of stars, gas and dark matter halos, with one million particles for each component. The model also accounts for energy dissipation in the interstellar gas, and the onset of star formation[22], reproducing both the global morphology of the NGC5291 system, and the formation of recycled dwarf galaxies in it. These simulations enable us to date the formation of the system and to study its three-dimensional morphology. According to our model[23], the ring formed during a galaxy collision 360 million years ago and is seen inclined by 45° from the line-of-sight (Figs. 1, S1, S2).

The rotation curve of a galaxy traces the rotational velocity of the disk as a function of radius and provides a direct measure of the total (visible and dark) mass within that radius. We use our VLA observations to derive the rotation curves of three recycled dwarf galaxies around NGC5291, and compare the mass inferred this way with their visible mass. The most luminous recycled galaxy in this system, NGC5291N, contains $5.7 \times 10^8$ $M_\odot$ (solar mass) of atomic gas, $2 \times 10^8$ $M_\odot$ of molecular gas traced by the emission of the CO molecule, and $1.1 \times 10^8$ $M_\odot$ of stars. We hence derive a total visible mass of $8.8 \pm 1 \times 10^8$ $M_\odot$ inside a radius of 3.7 kpc. The VLA data reveal a velocity gradient tracing the rotation of the object up to radii of ~4.5 kpc (Fig. 2 and S3). The rotational velocity, corrected for inclination as indicated by our model, is 70 km s$^{-1}$ at a radius of 3.2 kpc, implying that the mass actually present in the system – the so-called dynamical mass – is $30 \pm 8.6 \times 10^8$ $M_\odot$(SD). The error bar accounts for the noise in the data and various uncertainties entering the method for determining the mass[23]. This system then contains an unseen component, with about twice the visible mass. The visible mass is insufficient to explain not only the high rotational velocity but also the flat rotation curve: the rotational velocity remains constant beyond the radius at which the visible material is concentrated (Fig. 2).

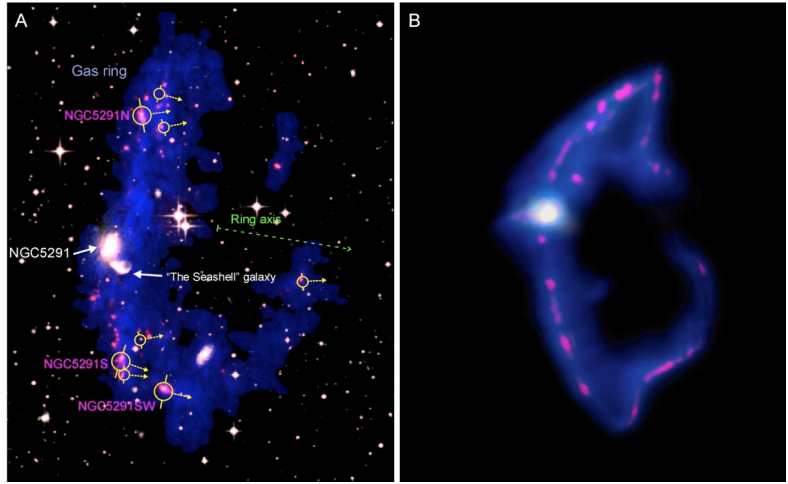

**Fig. 1.** Gas ring and recycled dwarf galaxies around NGC5291. (**A**) VLA atomic hydrogen 21-cm map (blue) superimposed on an optical image (white). The UV emission observed by GALEX (red) traces dense star-forming concentrations. The most massive of these objects are rotating with the projected spin axis as indicated by arrows. In the three most massive companions, we quantify the rotation and estimate their mass content. (**B**) Numerical simulation of a galaxy collision that has led to the formation of a similar system where the central galaxy is the progenitor of a large, asymmetrical and partial ring. Self-gravitating clumps have assembled within the ring to become new dwarf galaxies. The ring is seen projected at an angle of 45° from the line-of-sight. The recycled dwarf galaxies have individual rotation axes closely aligned with that of the large-scale ring axis, which indicates that they are also inclined by 45°.

In the second brightest object, NGC5291S, the visible mass is $9.3 \pm 1 \times 10^8$ $M_\odot$ whereas its dynamical mass is $27 \pm 8.5 \times 10^8$ $M_\odot$. This object must thus contain a dark component as well, also with about twice the mass of the visible matter. The data on the third most massive object (NGC5291SW) lead to a similar dark-to-visible mass ratio, with a visible mass of $5 \pm 1.5 \times 10^8$ $M_\odot$ and a dynamical mass of $12 \pm 4.5 \times 10^8$ $M_\odot$, albeit with larger uncertainties. The other recycled objects in this system are less massive, resulting in error bars that are too large for the purpose of constraining the dynamical mass.

It is possible to derive dynamical masses of galaxies from their rotational velocity even when they are barely resolved[24], provided that some assumptions are met, in particular the gas being in equilibrium and moving on quasi-circular orbits. It is therefore important to ascertain that the rotation curves of the recycled dwarfs are not affected by velocity anomalies, as a result of their young age or the presence of the massive progenitor.

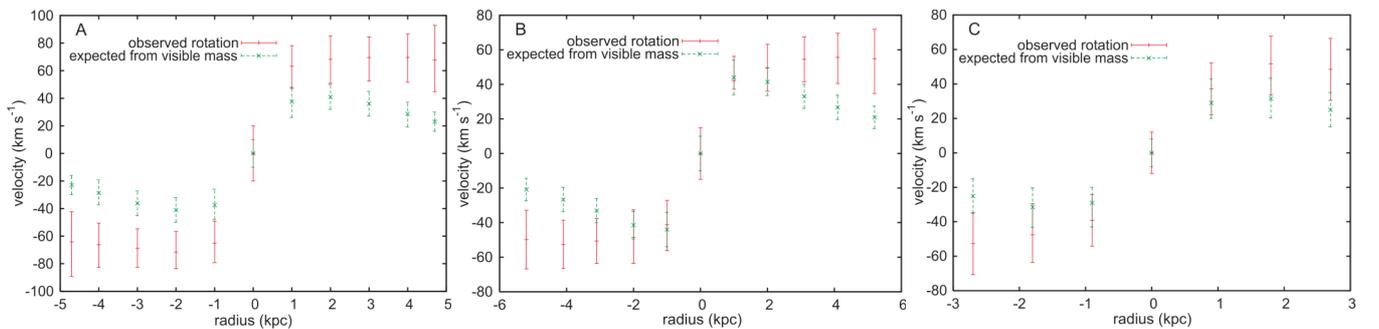

**Fig. 2.** Observed rotation curves (red) compared to what is expected for the visible mass alone (green) for the recycled galaxies NGC5291N (**A**), NGC5291S (**B**) and NGC5291SW (**C**). The rotation curves are derived from HI spectra, measuring the velocity where the emission has dropped to 50% of the peak level[33]. The error bars (SD) incorporate the effect of noise, the uncertainty inherent in the deprojection, as well as the uncertainty involved in converting the observed velocity gradients to intrinsic rotational velocity[23]. The green rotation curves show the predicted velocities for a disk with the visible mass of each object, observed at the same resolution. Associated error bars correspond to the uncertainty on the visible mass. With the visible mass alone, the velocities would be lower and drop at large radius, which is in contradiction with the flat rotation curves in the outer parts and confirms the presence of a "dark" component.

The symmetrical rotation pattern points already to the fact that these objects are rotationally supported and close to equilibrium. To further assess the validity of our results, the same analysis was applied to the dwarf galaxies formed in our numerical simulations. These simulations resolve the internal structure of the dwarf galaxies like spiral arms (Fig. S7), and indicate that they have reached a state of rotationally supported equilibrium (Fig. S4). Their kinematics was analysed at the same resolution as that of the observations, and we verified that their mass could be retrieved with the required accuracy. The large observed velocities cannot thus be caused by any velocity anomalies, and our analysis indicates the presence of a dark component at a confidence level of 98% for both NGC5291N and NGC5291S[25] and 95% for NGC5291SW (Fig. 3).

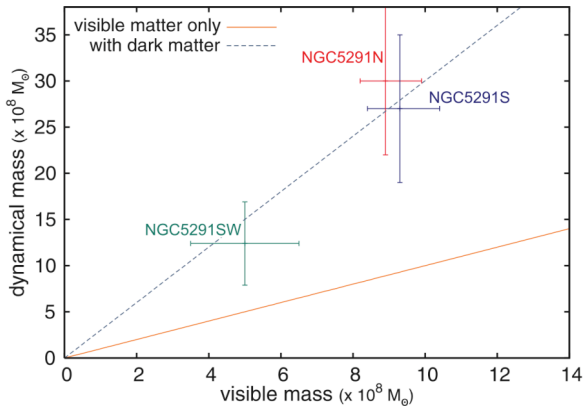

**Fig. 3.** Visible and dynamical masses of three recycled galaxies. Dark matter-free objects would follow the orange full line. The three systems show a similar trend with a dark mass about twice that of the visible mass inside the same radius. The error bars (SD) for both axes are indicated. The similar result obtained on three objects makes the overall confidence level*(25)* in the detection of dark matter greater than 99%.

That recycled dwarf galaxies contain a massive unseen component of about twice the visible mass is surprising, even if they do not contain as much dark mass as classical, first generation, dwarf galaxies*(2,6)* where dark-to-visible mass ratios can be as large as ten*(26)*. Within standard paradigms, recycled galaxies were expected to contain at most a few percent*(12)* of dark matter. This is corroborated by our simulations which were conducted within the standard framework with all dark matter in spheroidal halos, and none in the disks of spiral galaxies: the rotational velocities predicted this way in recycled dwarfs differ from the observed ones (Fig. S6, S7), which confirms that an additional dark component is required in the observed system. If this dark mass was non-baryonic, its properties would have to differ from the prevailing Cold Dark Matter (CDM) model. Only material initially in a rotating disk in the progenitor galaxy can participate in the creation of recycled dwarfs, whereas CDM resides in non-rotating spheroidal halos. Given the success of ΛCDM scenarios to explain the large-scale structure of the Universe*(27)*, we are inclined to consider that the unseen matter in recycled dwarf galaxies is more likely baryonic. This would imply that the "missing baryons" do not all reside in the warm-hot intergalactic medium: hot gas forms diffuse halos that do not provide material to the recycled galaxies. A significant fraction of dark baryons would then be located in a (potentially thick) disk. This cannot be in the form of low-luminosity old stars, that are known to be absent from the dwarf galaxies around NGC5291*(28)*. The most likely candidate is hydrogen in dense molecular form. The $H_2$ molecule cannot be directly observed but is usually traced by emission lines of the carbon monoxide (CO) molecule. The emission from CO can be converted to a molecular gas mass*(29)*, using a conversion factor that is expected to depend on the heavy element abundance (metallicity) of the gas*(30)*. Here we used the usual conversion factor derived in our Galaxy for gas with Solar metallicity. The metallicity of the recycled galaxies in NGC5291 is about half solar*(18)*, similar to that encountered in the outer regions of spiral galaxies where the conversion factor could change*(31,32)* by factors of typically 2, or at most 3. To account for the missing mass in the recycled galaxies, however, a change of at least a factor of ten would be needed. This means we would be dealing either with cool $H_2$, traced by the CO molecule but much less efficiently than generally assumed, or with a sizeable fraction of cold $H_2$ not traced at all by CO*(15)*.

Collisional debris from galaxies hence appears to contain twice as much unseen matter as visible matter. Although this result could be explained by a modification of Newtonian gravity, it more likely indicates that a significant amount of dark matter resides within the disks of spiral galaxies. The most natural candidate is molecular hydrogen in some hard to trace form. Further simulations including this form of dark matter and comparison with higher resolution observations of recycled galaxies will be required to directly constrain the exact properties of this unseen component.

independent variables.

34. The numerical simulations were carried out at CEA/CCRT and CNRS/IDRIS. Numerical models have beneficiated from input from the collaboration HORIZON, and we grateful to Françoise Combes and Romain Teyssier. Valuable comments on the dynamical analysis and/or general results from Jonathan Braine, Peter Weilbacher, Isabelle Grenier, Yves Revaz, François Boulanger, and François Hammer are gratefully acknowledged. We made use of data from the Digitized Sky Survey, produced at the Space Telescope Science Institute under U.S. Government grant NAG W-2166.

Full version with appendix (supporting online material) available at:

http://www.sciencemag.org/sciencexpress/recent.dtl

or

http://www.uni-sw.gwdg.de/~paduc/articles/bournaud_tdg_dm.pdf